\documentclass[pdflatex, sn-nature]{sn-jnl}

\usepackage{graphicx}%
\usepackage{multirow}%
\usepackage{amsmath,amssymb,amsfonts}%
\usepackage{amsthm}%
\usepackage{mathrsfs}%
\usepackage[title]{appendix}%
\usepackage{xcolor}%
\usepackage{textcomp}%
\usepackage{manyfoot}%
\usepackage{booktabs}%
\usepackage{algorithm}%
\usepackage{algorithmicx}%
\usepackage{algpseudocode}%
\usepackage{listings}%
\usepackage[normalem]{ulem}
\usepackage[section]{placeins}
\usepackage{lineno}
\DeclareUnicodeCharacter{2212}{-}

\begin{document}

\title[Article Title]{Nematicity of a Magnetic Helix}


\author[1,2]{R. Tumbleson}
\author[3]{S. A. Morley}
\author[2,4]{E. Hollingworth}
\author[2,5]{A. Singh}
\author[2]{T. Bayaraa}
\author[6,9]{N. G. Burdet}
\author[3]{A. U. Saleheen}
\author[3]{M. R. McCarter}
\author[1,2]{D. Raftrey}
\author[7]{R. J. Pandolfi}
\author[6,9]{V. Esposito}
\author[6]{G. L. Dakovski}
\author[6]{F.-J. Decker}
\author[6]{A. H. Reid}
\author[9]{T. A. Assefa}
\author[1,2]{P. Fischer}
\author[2,8]{S. M. Griffin}
\author[3]{S. D. Kevan}
\author[2,4,10]{F. Hellman}
\author[6,9]{J. J. Turner}
\author*[1,2,3]{S. Roy}

\affil[1]{\orgdiv{Department of Physics}, \orgname{University of California, Santa Cruz}, \orgaddress{\city{Santa Cruz}, \postcode{95064}, \state{California}, \country{USA}}}

\affil[2]{\orgdiv{Materials Sciences Division}, \orgname{Lawrence Berkeley National Laboratory}, \orgaddress{\city{Berkeley}, \postcode{94720}, \state{California}, \country{USA}}}

\affil[3]{\orgdiv{Advanced Light Source}, \orgname{Lawrence Berkeley National Laboratory}, \orgaddress{\city{Berkeley}, \postcode{94720}, \state{California}, \country{USA}}}

\affil[4]{\orgdiv{Department of Physics}, \orgname{University of California, Berkeley}, \orgaddress{\city{Berkeley}, \postcode{94720}, \state{California}, \country{USA}}}

\affil[5]{\orgdiv{Materials Science Division}, \orgname{Argonne National Laboratory}, \orgaddress{\city{Lemont}, \postcode{60439}, \state{Illinois}, \country{USA}}}

\affil[6]{\orgdiv{Linac Coherent Light Source}, \orgname{SLAC National Accelerator Laboratory}, \orgaddress{\city{Menlo Park}, \postcode{94025}, \state{California}, \country{USA}}}

\affil[7]{\orgdiv{Center for Advanced Mathematics for Energy Research Applications, Applied Mathematics and Computational Research Division}, \orgname{Lawrence Berkeley National Laboratory}, \orgaddress{\city{Berkeley}, \postcode{94720}, \state{California}, \country{USA}}}

\affil[8]{\orgdiv{Molecular Foundry}, \orgname{Lawrence Berkeley National Laboratory}, \orgaddress{\city{Berkeley}, \postcode{94720}, \state{California}, \country{USA}}}

\affil[9]{\orgdiv{Stanford Institute for Materials and Energy Sciences}, \orgname{ SLAC National Accelerator Laboratory and
Stanford University}, \orgaddress{\city{Menlo Park}, \postcode{94025}, \state{California}, \country{USA}}}

\affil[10]{\orgdiv{Department of Materials Science and Engineering}, \orgname{University of California, Berkeley}, \orgaddress{\city{Berkeley}, \postcode{94720}, \state{California}, \country{USA}}}



\abstract{\textbf{A system that possesses translational symmetry but breaks orientational symmetry is known as a nematic phase. While there are many examples of nematic phases in a wide range of contexts, such as in liquid crystals, complex oxides, and superconductors, of particular interest is the magnetic analogue, where the spin, charge, and orbital degrees of freedom of the electron are intertwined. The difficulty of spin nematics is the unambiguous realization and characterization of the phase. Here we present an entirely new type of magnetic nematic phase, which replaces the basis of individual spins with magnetic helices. The helical basis allows for the direct measurement of the order parameters with soft X-ray scattering and a thorough characterization of the nematic phase and its thermodynamic transitions. We discover two distinct nematic phases with unique spatio-temporal correlation signatures. Using coherent X-ray methods, we find that near the phase boundary between the two nematic phases, fluctuations coexist on the timescale of both seconds and sub-nanoseconds. Additionally, we have determined that the fluctuations occur simultaneously with a reorientation of the magnetic helices, indicating that there is spontaneous symmetry breaking and new degrees of freedom become available. Our results provide a novel framework for characterizing exotic phases and the phenomena presented can be mapped onto a broad class of physical systems.
}}




\maketitle


Nematic order describes a system that possesses orientational order but lacks translational order \cite{Dierking2003}. While its origin lies in the research field of liquid crystals, the phenomena associated with the nematic phase has turned out to be quite general with broad applicability in a wide range of contexts, including bacterial suspensions \cite{Doostmohammadi2018}, active matter \cite{Mur2022}, polymer networks \cite{Ozenda2020}, and superconductivity \cite{Nie2022, Roßler2022}, to name a few. A nematic phase is an example of a layered system where the phase transitions of the layers are mediated by unbinding of topological defect pairs \cite{Toner1981}. Specifically, electronic nematicity in quantum materials is considered to be a new state of matter and is due to low dimensional electron correlation effects \cite{BorziScience}. These effects manifest as anisotropic transport in superconductors \cite{Fradkin2007}, charge density wave driven nematic phase \cite{NematicCDW}, or magnetoresistance in complex oxides \cite{Cooper2002, BorziScience}. Further, due to the  presence of competing intertwined phases with nearly degenerate ground states, often in the region of phase space where nematicity is observed there exists other exotic phases, suggesting that the phases are not independent of one another. Therefore, understanding the phase transitions of nematics not only gives insight into the phase itself, but is also illuminating to other nearby hidden phases.


The way in which phases and their transitions are characterized is by their order parameter(s). The most straight forward example is a magnetic system where the global magnetization is taken as the order parameter, and it describes the ensemble of local magnetic moments present. The observation of nematic order in magnetic systems is of great scientific interest since the electronic degrees of freedom, i.e. spin, charge, and orbital, are coupled and there are fundamental differences compared to other nematic phases, such as a potential ferromagnetic background, which supplants entirely new boundary conditions on the problem. While recently there have been a few experimental observations of spin nematics \cite{BJKimNature, Kohama2019, Moon2023}, the fundamental constituents have been restricted to individual spins at well-defined lattice sites. This, however, is not the only magnetic order that can exist, and in this article, we present an entirely new nematic phase based on the emergent order associated with a magnetic helical phase.

Magnetic spin textures offer a fertile playground for studying topological phases and their transitions, offering a diverse selection of different phases, experimentally measurable length-scales, and a coupling to external stimuli. A canonical example is the B20 crystalline polymorph of FeGe, which has seen considerable interest over the past decade due to the discovery that it can host a Skyrmion crystal close to room temperature \cite{Yu2010}, stabilized by the inversion symmetry breaking of the atomic unit cell. More recently, however, it has been found that a well-defined crystal structure is not a requirement to host magnetic spin textures \cite{Streubel2021}. This opens the possibility to tailor different phases by controlling the atomic composition and measure phase transitions with reduced constraints to the underlying crystalline environment \cite{Bouma2020, Singh2023}.

To thoroughly understand the thermodynamic transitions of the nematic phase in a magnetic system, we performed static and dynamic resonant soft X-ray scattering (RSXS) and X-ray Photon Correlation Spectroscopy (XPCS) measurements on amorphous Fe$_{51}$Ge$_{49}$ (\textit{a}-Fe$_{51}$Ge$_{49}$) films. Additionally, we combined micromagnetic simulations and density functional theory (DFT) calculations with our experimental findings to further investigate the degrees of freedom of the order parameters. We found that the helical magnetic texture that stabilizes in \textit{a}-Fe$_{51}$Ge$_{49}$ can be described as chiral nematic (N*) with two distinct thermodynamic phases exhibiting characteristic spatio-temporal correlations. We identify the two phases by their correlation length with a low temperature phase, $\rm N^*_{\rm \xi_>}$, possessing a larger correlation length and a high temperature phase, $\rm N^*_{\rm \xi_<}$, possessing a shorter correlation length. Just before the $\rm N^*_{\rm \xi_<}$ phase, we found that the system becomes unstable, and fluctuations coexist on the timescale of both seconds and sub-nanoseconds. Finally, we found that the onset of the fluctuations coincides with a reorientation of the helical propagation direction from being purely in-plane to canted out-of-plane, resulting in an availability of new degrees of freedom; the cant angle of the helix propagation vector. This indicates that there is crossover in the dimensionality of the degrees of freedom in the system.

\begin{figure*}[htbp]
\centering
\includegraphics[width=\linewidth]{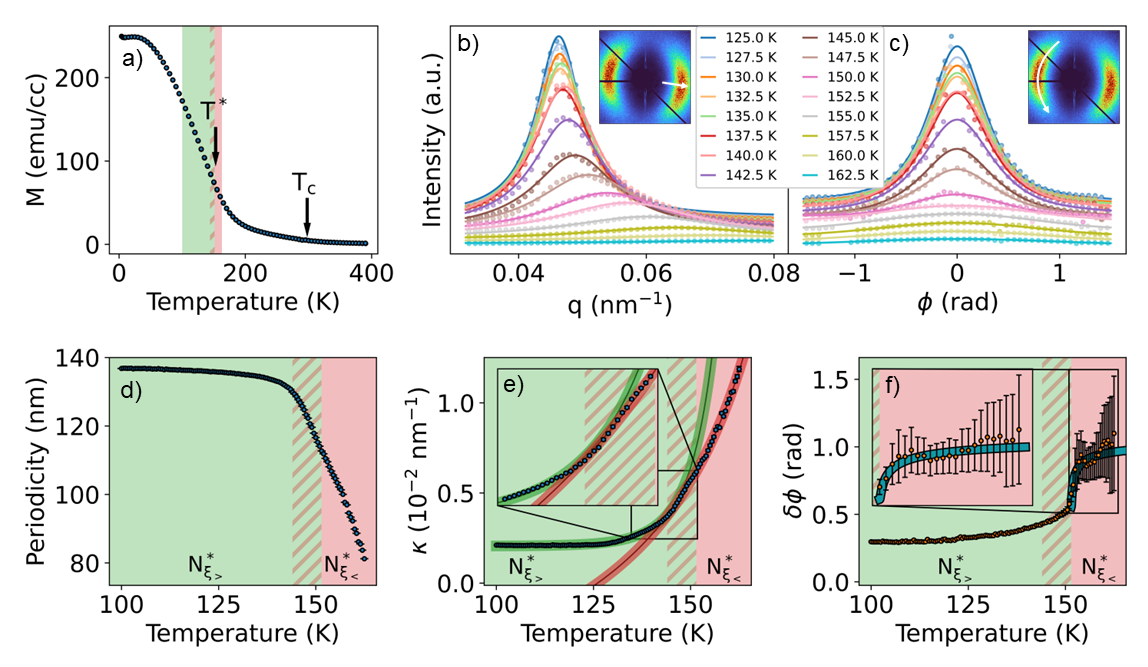}
\caption{Static scattering of \textit{a}-Fe$_{51}$Ge$_{49}$. (a) Magnetization versus temperature with an applied in-plane field of 1000 Oe. The Curie temperature ($\rm T_C$) was found to be 298~K from analysis of M(H, T) and $\chi$(T) up to 400~K. The $\rm N^*_{\rm \xi_>}$ to $\rm N^*_{\rm \xi_<}$ transition temperature ($\rm T^*$) is $\sim$151.5~K, determined by the fit in (f). (b)-(c) Quantification of the static scattering patterns. Exemplary averages are between the temperature range of 125~K (top) and 162.5~K (bottom). The white arrow in the insets show the averaging directions. Solid lines correspond to (b) Lorentzian and (c) pseudo-Voigt fits to the data for the radial-q and azimuthal-$\phi$ averages, respectively. (d) Average helix periodicity extracted from the peak position in (b). (e) The inverse correlation length, $\kappa$ extracted from the full-width at half maximum (FWHM) in (b) showing the translational order parameter and a crossover from power law (green line) to exponential (red line) functional form. (f) The FWHM, $\delta \phi$, of the azimuthal distribution in (c) showing the orientational order parameter. Solid line in (f) shows a fit to Eq. \ref{eq:nematic}.}
\label{fig:static_scattering}
\end{figure*}

 
 Within this article, we refer to the phases using standard nomenclature used to describe liquid crystals and has broadly been extended to other physical systems. We would like to point out, however, that the magnetic helices do not preserve inversion symmetry, as is the case for rod-like directors. Despite this distinction, all elastic properties are identical to more traditional nematics \cite{Ostlund1981}. Additionally, an * is used to emphasize that the magnetic helix is a chiral entity. The temperature dependence of the magnetization for \textit{a}-Fe$_{51}$Ge$_{49}$ is shown in Fig. \ref{fig:static_scattering} (a) where the highlighted regions correspond to the temperature range measured by RSXS in this study, $\rm T^*$ marks the $\rm N^*_{\rm \xi_>}$-$\rm N^*_{\rm \xi_<}$ transition temperature, and $\rm T_{\rm C}$ is the Curie temperature. In Fig. \ref{fig:static_scattering} (b) and (c) we show temperature dependence of the radial and azimuthal averages from the recorded scattering patterns (insets) and corresponding fits to Lorentzian and pseudo-Voigt functions, respectively (solid lines are fits). The average  helical periodicity of the system shows an interesting temperature dependence, as shown in Fig. \ref{fig:static_scattering} (d). At lower temperatures, the periodicity is stable with a value around 135~nm. When the temperature is increased above $\sim$144~K, there is a turning point and the periodicity decreases more rapidly. This decrease in periodicity is linear with increasing temperature until the signal is no longer observed at T $\sim$ 163~K. It is important to note that this disappearance of the X-ray scattering is $\sim$100~K below $\rm T_{\rm C}$, where there is an appreciable in-plane net magnetization, as seen in Fig. \ref{fig:static_scattering} (a). This result indicates that for T* $\le$ T $\le$ $\rm T_{\rm C}$ the magnetic helical texture either orients itself in such a way that the periodic structure is not observable in transmission scattering or regions between the magnetic helices are weakly ferromagnetic.

 To quantify the positional order of the system, the inverse correlation length, $\kappa$, was extracted (Fig. \ref{fig:static_scattering} (e)) from the full-width at half-maximum (FWHM) of the radial averages. Below the periodicity inflection point, $\kappa$ grows algebraically with an initial correlation length of $\sim$500~nm. At the inflection point, there is a crossover from an algebraic scaling to an exponential. The different functional scaling with temperature of the correlation length within different temperature windows indicates a multi-step order-disorder transition.
 
From the azimuthal averages (Fig. \ref{fig:static_scattering} (c)), we obtain the orientational order by determining the angular spread $\delta \phi$, or FWHM of the distribution shown in Fig. \ref{fig:static_scattering} (f). In contrast to the translational order, $\delta \phi$ shows a slow increase until $\sim$150~K. At this point, $\delta \phi$ sharply increases and then plateaus around 155~K. The solid line in \ref{fig:static_scattering} (f) is a fit to the following expression

\begin{equation}
    \delta \phi \propto e^{- \rm T_R^{-1/2}}
    \label{eq:nematic}
\end{equation}

where $\rm T_R \propto \rm T - \rm T^*$. This functional scaling of the broadening of the azimuthal X-ray scattering peak is predicted \cite{Ostlund1981} near a nematic phase boundary, suggesting that the $\rm N^*_{\rm \xi_<}$ phase maintains nematic character. The fit to \ref{eq:nematic} pins our transition temperature, $\rm T^*$ to $\rm T^*=$ 151.5~K. We note here once again the important distinction that we are relating the distribution of magnetic spin helices to that of the directrix in a conventional nematic. The distinct changes in the azimuthal data while translation correlation being very short confirms that the system has entered into a new phase, $\rm N^*_{\rm \xi_<}$, where the orientational order is destroyed slightly, as evidenced by the peak broadening, but is still preserved, since the two-fold symmetric scattering pattern is preserved.

\begin{figure}[htbp]
\centering
\includegraphics[width=\linewidth]{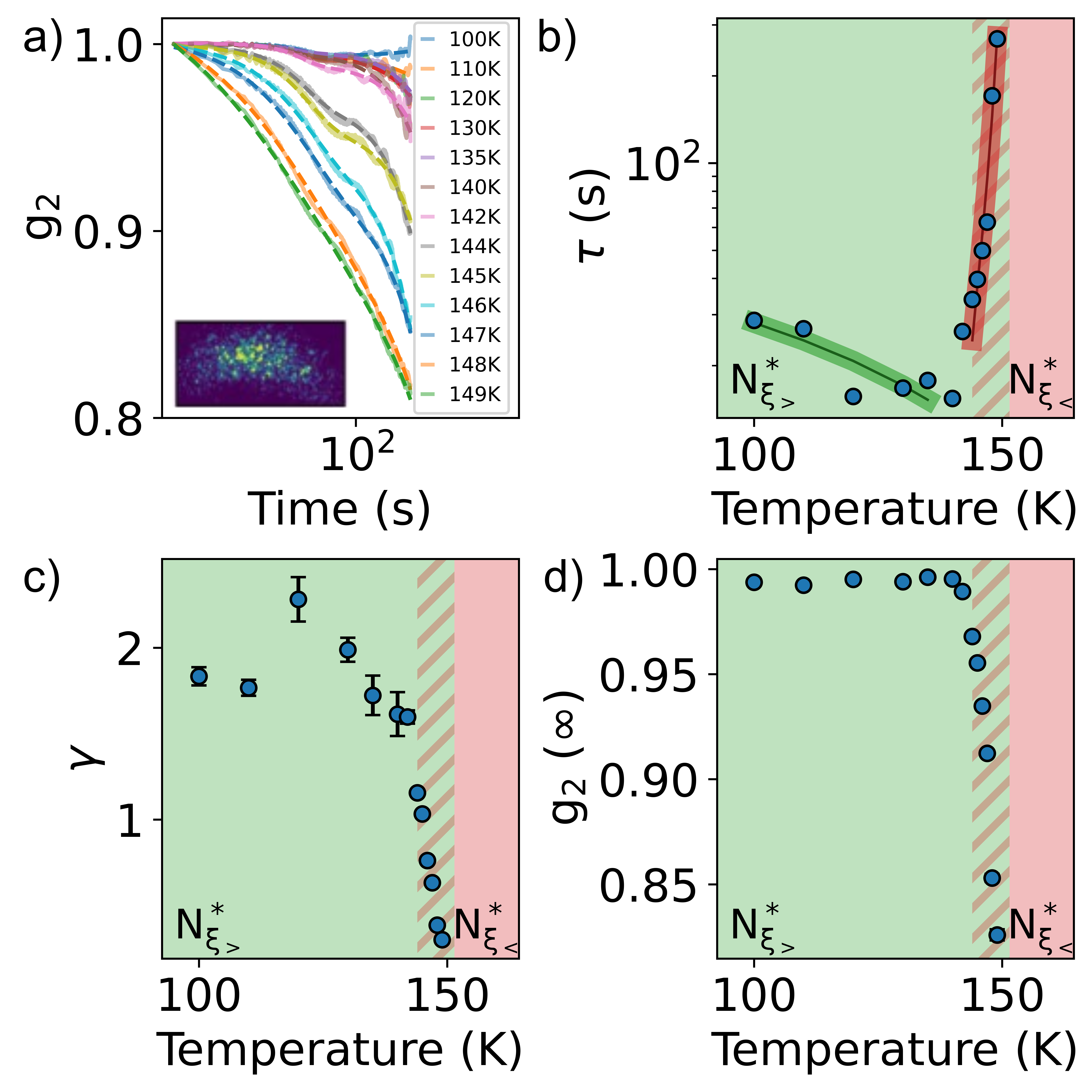}
\caption{Dynamic correlations of \textit{a}-Fe$_{51}$Ge$_{49}$. (a) X-ray photon correlation spectroscopy (XPCS) of a coherent scattering pattern (inset) showing complex dynamics with multiple relevant time-scales. The solid lines are calculated autocorrelation values, and the dashed lines are fits to Eq. \ref{eq:mod_kww}. The characteristic relaxation time of the exponential term (b), the stretching exponent (c), and the constant offset (d) extracted from (a) as a function of temperature from 100~K to 149~K measured on heating after cooling in zero field. The green and red lines in (b) correspond to linear and exponential trends, respectively, and are guides to the eye.}
\label{fig:xpcs}
\end{figure}

In addition to these static scattering signatures of the phase transition, unique temporal correlations are also present upon approaching the $\rm N^*_{\rm \xi_<}$ phase, as evidenced by the time dynamics studies discussed presently. Using the technique of X-ray Photon Correlation Spectroscopy (XPCS)\cite{sinha2014x,Lehmkuhler2021}, we observe a change from a small spatial volume of fluctuations at low temperatures to an increasing volume of fluctuations at an overall slower timescale as the $\rm N^*_{\rm \xi_<}$ phase is approached from below. A typical scattering pattern consisting of speckles is shown in the inset of Fig. \ref{fig:xpcs} (a) and solid lines show experimental $g_2$ autocorrelation values. Note that all of the $g_2$ data is preprocessed by subtracting unity and dividing by the first $g_2$ value. In our $g_2$ autocorrelation calculation, a region containing the entire speckle pattern of a magnetic peak is chosen so the dynamics measured are a statistical ensemble of the fluctuations present at the length-scale of the helical periodicities. It is clear from our $g_2$ data that there is not a single characteristic timescale representative of the system. Each temperature shows dynamics at a faster timescale with decorrelations on the order of seconds followed by dynamics on a much slower timescale. 

These multi-timescale dynamics exclude the possibility of analysis by the common treatment of fitting a single stretched exponential function of Kohlrausch-Williams-Watts (KWW) form \cite{VincePRL, Lehmkuhler2021}. In addition to this complicated temporal structure, the timescale of the slower dynamics are prohibitively long to adequately sample experimentally and fit the data to a sum of multiple exponential functions. Instead, we fit our data with a single stretched exponential function modified by a linear offset, given by the equation:

\begin{equation}
    g_2 \propto e^{-(t/\tau)^{\gamma}} + mt + g_2(\infty)
    \label{eq:mod_kww}
\end{equation}

Here, $\tau$ is the relaxation time, $\gamma$ is the stretching exponent, and $g_2(\infty)$ is a constant. This modified exponential includes only a single additional variable, m, with linear time dependence which provides a simple fit to allow long timescale fluctuations and enables the possibility of extracting the fast time-scale dynamics. Fits to Eq. \ref{eq:mod_kww} are shown by the dashed lines in Fig. \ref{fig:xpcs} (a). Extracted fit parameters are given in Fig. \ref{fig:xpcs} (b)-(d). The relaxation time $\tau$ describes the characteristic timescale of the fast fluctuations, the stretching exponent represents the type of dynamics present, and $g_2(\infty)$ is a measure of the spatial volume of fluctuations over the measurement time. 

Our analysis indicates that for T$\sim$144~K all of the extracted parameters show a change in slope, indicating that this temperature is where the ground state becomes unstable. In the low temperature region (T $<$ 144~K), the characteristic relaxation time exhibits a relatively fast relaxation rate that is on the order of seconds. This coincides with a stretching exponent around 1.5, typical for systems with jammed dynamics \cite{VincePRL,Cipelletti2003}, and a $g_2(\infty)$ value close to one. This would indicate that the system is mostly static with only a small portion of the sample fluctuating, consistent with reported entropically-driven small movements of topological defects in the helical phase of crystalline B20 FeGe and its dynamics well below the helical ordering temperature \cite{Dussaux}.

As the temperature is increased above 144~K, which is the onset of the instability region, the characteristic faster fluctuations rapidly slow down. The stretching exponent goes from faster than exponential ($\gamma >$ 1) to slower than exponential ($\gamma <$ 1) and a sharp decrease in $g_2(\infty)$ is observed. This result, although unintuitive, means that on average a slower timescale is present. With increasing temperature our signal is increasingly dominated by collective slower motion that covers a larger sample volume. Thus, even though faster fluctuations due to topological defects are still present, a majority of the fluctuations being measured are slower and that overshadows the faster fluctuations in the $g_2$ calculation. This is further bolstered by our nanosecond fluctuation measurements to be discussed below. 

To get further insight into the fluctuations at the nanosecond timescale in the region of instability, a two-pulse based X-ray correlation experiment was performed \cite{SeabergPRL, Shen2021, Seaberg2021} in which the speckle visibility of two integrated images is calculated on an individual photon level (see supplementary material for additional details) \cite{Goodman2015}. The speckle contrast extracted from a two-pulse measurement is normalized by a single-pulse measurement such that a contrast value close to one is a static measurement. Contrast deviations below one therefore represent a lower visibility and measure fluctuations on the timescale probed by the two pulses.

\begin{figure}[htbp]
\centering
\includegraphics[width=\linewidth]{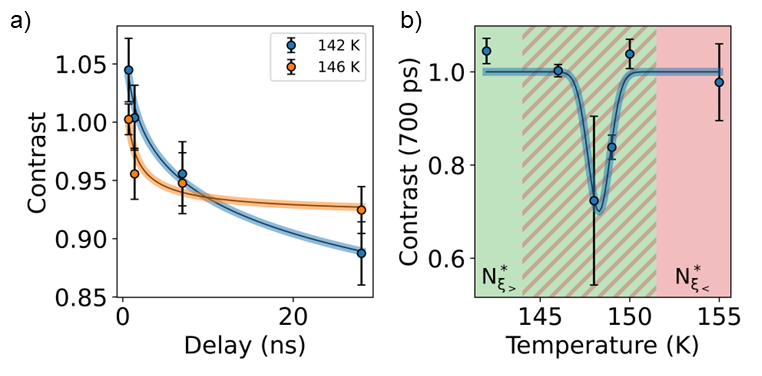}
\caption{Sub-nanosecond fluctuations near the phase transition at $\rm T^*$= 151.5~K. The contrast value was extracted from the probability distribution of photon-arrival events at a particular two-pulse delay time. (a) Just below (blue) and above (orange) the onset of the region of instability distinct decays of nanosecond dynamics are present. (b) While measuring with a fixed delay time of 700 ps, the enhancement of magnetic fluctuations are observed just before T* is reached. The solid line is a guide to the eye.}
\label{fig:lcls}
\end{figure}

Just below and above the temperature where the slow timescale fluctuations set in, the contrast was measured between 700~ps and 28~ns (Fig \ref{fig:lcls} (a)). It is evident that there exists nanosecond fluctuations with distinct decorrelation signatures, i.e. characteristic time and degree of decorrelation. More interestingly, a measurement at a single timescale of 700 ps as a function of increasing temperature (Fig. \ref{fig:lcls} (b)) through the region of instability shows a large enhancement of fluctuations just before the $\rm N^*_{\rm \xi_<}$ phase is reached. Increasing the temperature even further through $\rm T^*$ brings the contrast back to a value close to one, a result of noise dominating the signal since the peak intensity is decreasing from both the decreasing strength of the order and the faster fluctuations present as $\rm T^*$ is approached.

\begin{figure}[htbp]
\centering
\includegraphics[width=\linewidth]{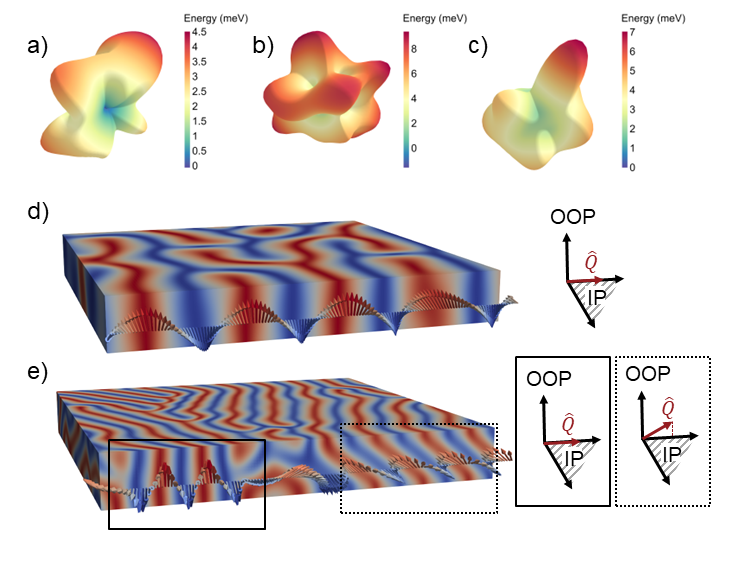}
\caption{Computational support of helical reorientaion. (a)-(c) Magneto-structural anisotropy energy surfaces are plotted for three amorphous structures prepared as described in methods. (d) Micromagnetic simulation showing the low-temperature phase where the helical propagation direction (Q-vector) is purely in-plane . (e) Simulation in the same region after the reorientation begins showing a decrease in periodicity and an inhomogeneous rotation of the Q-vector to be both in-plane (solid box) and canted out-of-plane (dashed box). Rendered regions in (d) and (e) represent a volume of $\sim$500 nm x 1 um x 80 nm.}
\label{fig:mumax}
\end{figure}

To complement our experimental observations we calculate, as described in methods, the magneto-structural anisotropy energy (MSAE) surfaces of three amorphous FeGe structures and report them in Fig. \ref{fig:mumax} (a)-(c). This calculation is the amorphous analogue of magneto-crystalline anisotropy energy in crystalline systems and is vital in determining the orientation of magnetic helices \cite{Preißinger2021}. MSAE is found to reach as high as 9.42 meV and the average highest MSAE for all three structures is 6.92 meV which is equivalent to a temperature of 80~K. This indicates that thermally induced switching of the magnetization axis could occur beyond 80~K which in line with our experimental observation of 144~K where the periodicity has an inflection point and the fluctuations set in. This type of thermal reorientation of the helical propagation direction is well-known to exist in the B20 crystalline form of FeGe where the helix switches from the (111) orientation to (100) at 279~K \cite{Lebecht1989}. In contrast, we believe that the reorientation in our system occurs over a broad temperature range and the helical propagation direction (Q-vector) is not a coherent reorientation, as evidenced by our micromagnetic simulations \ref{fig:mumax} (d)-(e) discussed below.

To simulate the effect of temperature on the magnetic structure using micromagnetic simulations, we varied the ratio of the exchange stiffness to the Dzyloshinskii-Moriya interaction (DMI) and matched the periodicity of the relaxed state to experimental scattering patterns (see supplementary materials). It is important to note that raising temperature alone in the simulations was not sufficient to reproduce the periodicity change observed in our experimental data. In the simulations, when exchange stiffness is reduced while keeping DMI constant (equivalent to raising temperature), regions with in-plane (solid box in Fig. \ref{fig:mumax} (e)) and others with canted out-of-plane (dashed box in Fig. \ref{fig:mumax} (e)) Q-vectors are observed, indicating an inhomogeneous rotation of the Q-vector. This inhomogeneity is a manifestation of the topological constraint of the system and results in the nucleation of defects in the 80~nm thickness (out-of-plane direction) of the sample. An example of such a defect is shown in Fig. \ref{fig:mumax} (e) between the two highlighted regions with different Q-vectors. This reorientation would also reduce the out-of-plane magnetic contrast when averaged along the film thickness despite retaining a helical spin structure, as observed in the scattering data.


Our study has identified a magnetic system with chiral nematic (N*) character where the role of the director is replaced by magnetic spin helices. The thermodynamic order-disorder transition of N* can be distinguished into two distinct phases with a low temperature $\rm N^*_{\rm \xi_>}$ phase continuously transforming into a high temperature $\rm N^*_{\rm \xi_<}$ phase with distinct spatio-temporal correlations. The $\rm N^*_{\rm \xi_>}$ phase is characterized by a well-ordered helical orientation and pitch with dynamics consisting of only a small spatial volume of the sample. Despite having a well-ordered Q-vector, this phase should be distinguished from a smectic phase, where long-range one-dimensional order exists. The radial broadness of the scattering peak is well above the resolution limit of the measurement, meaning that there is an appreciable amount of defects present even at the lowest temperatures measured.

When the temperature is raised to just below the transition temperature T*, there is a turning point in the helical periodicity, a crossover in the functional scaling of the inverse correlation length, and a striking change in the dynamics that are present. This marks a region of instability between the phases (identified by the hatched region in Figs. \ref{fig:static_scattering}-\ref{fig:lcls}). As illuminated by the micromagnetic simulations and DFT calculations, there is a crossover in the dimensionality of the system. At low temperatures, the thermal energy is only strong enough to displace the Q-vector direction within a plane, evidenced by the relatively sharp Lorentzian profile of the radial scattering peaks. At higher temperatures, however, there is more freedom to change the helical pitch, since the radial scattering peak broadens and the average helical periodicity changes. In addition, the Q-vector loses the planar orientational confinement, since the azimuthal scattering peak broadens more rapidly with temperature and the calculations show that the helices can orient in three dimensional space. These changes show that there is spontaneous symmetry breaking and a crossover in the degrees of freedom is observed, allowing for the rich dynamics measured.

Deep in the $\rm N^*_{\rm \xi_>}$ phase, the topological defects introduce several local minima in the energy landscape. There are only a few configurations that are accessible via thermal excitations on an experimentally feasible timescale. The dynamics that we measured are likely the motion of individual topological defects, such as dislocations that possess a high energy density or localized regions of a domain wall that are not in equilibrium even at the lowest temperatures \cite{Dussaux, Schoenherr2021}; the motion is heavily constrained. Superposed on these localized dynamics is the equilibrium global motion of the helix. These multi-time and space scale events continue and merge towards a coherent event as the $\rm N^*_{\rm \xi_<}$ phase is approached.

There are interesting questions that arise on the role of shape anisotropy on nematicity. We observe that the last measured helical periodicity before the scattering is lost coincides with both the correlation length and the thickness of the film (80 nm). Additional experiments where the film thickness is varied could reveal some new critical scaling and potentially enable the stabilization of different exotic phases. This can be combined with the tuning of different interaction strengths, since as the Fe concentration is increased so does the order of the system \cite{Singh2023}. This could enable measurements that couple to external stimuli, such as quantum oscillations in an applied magnetic field, of critical points near smectic, nematic, and isotropic phase boundaries in a Fermionic liquid-crystal type system. 

In connection with other nematic phases, the results presented in this article can find relevance to many other systems. For example, it has been long predicted that a chiral nematic phase can disorder by breaking apart into groups of chiral domains of size dictated by the density of dislocation defects \cite{Toner1981}. In superconductors, nematic phases can be observed in both the normal and superconducting phases with open questions about how the superconducting and nematic order parameters couple or compete with each other \cite{Li2017,NematicCDW}.  It is conceivable that our discovered exotic nematic phase with a magnetic helix basis may have interesting optical and transport properties with potential applications in microelectronics and spintronics. 

Identification and investigation of nematicity in quantum materials is challenging. In particular, dynamical measurements that characterize the fluctuations are essential in understanding phase transitions \cite{JanoschekFluctuation}. In the case of superconductivity, the identification of the nematic phase is reliant on electron transport. While transport measurements are highly sensitive, spatio-temporal distributions of the order parameter(s) are inaccessible. This warrants the need for nematic systems that can be measured with coherent X-ray scattering since those distributions become available and can be combined with other spectroscopic measurements to provide multimodal characterization. In addition to this new sensitivity, with the increasing coherent flux in the upcoming diffraction limited light sources, these measurements are becoming more precise. This allows for further probing into the nematic phase and accessibility to faster timescales which will provide deep insight into the interplay of fluctuation and phases.

\section*{Methods}

Films of \textit{a}-Fe$_{51}$Ge$_{49}$ were grown on Si$_3$N$_4$ membranes by DC magnetron co-sputtering as previously reported \cite{Singh2023}. Static and dynamic RSXS measurements were performed at Cosmic-Scattering beamline 7.0.1.1 at the Advanced Light Source. A transmission scattering geometry was employed wherein a linearly horizontal polarized X-ray beam tuned to the Fe L$_3$ edge was incident on the sample. This measurement is strictly sensitive to the out of plane component of the magnetization. The scattered signal was collected by a LBNL Fast Charge Coupled Device (CCD) camera placed 284~mm downstream of the sample \cite{pandolfi2018xi}. For the static measurements the sample was first cooled to 100~K, well below the helical ordering temperature, and subsequently heated until the scattering signal disappeared. For dynamic measurements we employed X-ray Photon Correlation Spectroscopy (XPCS). The exposure time was fixed at 200 ms for all measurements with a CCD readout time of 53~ms. A 7 $\mu$m pinhole was used to define a coherent incident beam and placed upstream of the sample. To probe faster timescale fluctuations, we performed a two-pulse X-ray correlation \cite{Shen2021} experiment at the ChemRIXS endstation \cite{Plumley2023} at the Linearly Coherent Light Source (LCLS) at Stanford Linear Accelerator Lab (SLAC). The energy was once again set to the Fe L$_3$ edge in transmission geometry and the pulse separation was varied between 700~ps and 28~ns. The speckle contrast was extracted from a statistical ensemble of sparse CCD images with a fixed pulse separation. Preliminary filtering of the data ensured that the two pulses were within a 20\% tolerance and all contrast measurements were normalized by a single pulse incident on the sample to eliminate the possibility of measuring dynamics that are not inherent to the sample. Sample dynamics were then extracted by tracking the speckle contrast change at different temperatures in the transition.

Micromagnetic simulations were performed using MuMax3 \cite{Vansteenkiste2014} with a volume of 512 x 512 x 20 cells and a cell size of 4~nm with in-plane periodic boundary conditions. To model the magnetic structure at different temperatures, a starting state was obtained by initializing with a random magnetization and subsequently minimizing the global energy. This state was then relaxed with a reduced exchange stiffness since as the transition temperature is approached from below, the exchange stiffness weakens \cite{Strelkov2020} and is the predominant energy term for the stabilization of the magnetic helices. 

To obtain representative amorphous structures, we used \textit{ab-initio} molecular dynamics (AIMD) simulations with the NVT ensemble as implemented in Vienna \textit{Ab-Initio} Software Package (VASP) \cite{Kresse1999} with the projector augmented-wave potentials \cite{Bloch1967}. We used the ‘melt-quench’ methodology to generate amorphous snapshots which was previously demonstrated on several systems \cite{Bouma2020}, and found to agree with experimental measurements of the structural and local electronic properties\cite{bayaraa2023,corbae,harrelson2021,cheng2020,sivonxay2020,sivonxay2022}. We constructed a cell of 96 atoms (Fe$_{0.5}$Ge$_{0.5}$) which were first randomly distributed in a cubic simulation cell using Packmol~\cite{packmol}. We obtained the stable liquid phase by equilibrating the pressure through a series of AIMD simulations at 3000~K, concurrently rescaling the unit cell between each AIMD simulation until it reached an internal pressure of 0 bar. Following this, we collected three snapshots at regular intervals after performing a 10 ps production run. We quenched these three snapshots following a stepped cooling scheme with 400 fs cooling and 1 ps isothermal steps. Final structures were optimized using VASP with an energy cutoff of 860 eV and $\Gamma$ point only sampling for the Brillouin Zone. All structural relaxations were performed until the Hellmann-Feynman force on each atom was less than 0.002 eV/\AA\ and included spin-orbit coupling (SOC) self consistently as implemented in VASP. We used the local-density approximation (LDA) \cite{LDA} for the final relaxations as this gave better agreement with the experimentally reported magnetic moments of 0.75 $\mu_B$ \cite{Streubel2021,bayaraa2023}. We calculated the magneto-structural anisotropy energy (MSAE) surfaces by varying the spin quantization axes over 50 points with SOC and comparing the total energies with respect to the lowest total energy.

\backmatter

\bmhead{Acknowledgments}
This work was supported by the U.S. Department of Energy, Office of Science, Office of Basic Energy Sciences, Materials Sciences and Engineering Division under Contract No. DE-AC02-05-CH11231 (NEMM program MSMAG). AUS acknowledges support from SUFD, DOE through award RoyTimepixDetector. MM acknowledges support from the LDRD project. Work at the ALS, LBNL was supported by the Director, Office of Science, Office of Basic Energy Sciences, of the US DOE (Contract No. DE-AC02-05CH11231). Computational resources were provided by the National Energy Research Scientific Computing Center and the Molecular Foundry, DOE Office of Science User Facilities supported by the Office of Science, U.S. Department of Energy under Contract No. DEAC02-05CH11231. The work performed at the Molecular Foundry was supported by the Office of Science, Office of Basic Energy Sciences, of the U.S. Department of Energy under the same contract. Data acquisition and visualization at beamline 7.0.1.1 was performed with Xi-CAM, a project supported by CAMERA, jointly funded by The Office of Advanced Scientific Research (ASCR) and the Office of Basic Energy Sciences (BES) within the DOE’s Office of Science. Portions of this work, both through the Materials Sciences and Engineering Division, as well as the use of the Linac Coherent Light Source, were supported by the U.S. Department of Energy, Office of Science, Basic Energy Sciences, under Contract No. DE-AC02-76SF00515. J.J.T. acknowledges support from the U.S. DOE, Office of Science, Basic Energy Sciences through the Early Career Research Program.

\section*{Declarations}

\subsection*{Conflict of Interest}
The authors declare no competing interests.

\subsection*{Availability of data and materials}
All data presented are available upon request.


\section*{Figures}\label{sec6}
\bigskip


\bibliography{bibliography}

\end{document}